\documentclass[aps,prl,twocolumn, amsmath,amssymb,superscriptaddress]{revtex4-1}
\usepackage{graphicx,color,array}
\usepackage{subfigure}
\usepackage{graphicx}
\usepackage{amsmath}
\usepackage{amsfonts}
\usepackage{amssymb}
\usepackage{latexsym,array,delarray,amsthm,epsfig}
\usepackage{multirow}
\usepackage{fancyhdr}
\usepackage{cancel}
\usepackage{paralist}

\newcommand{\Losc}{L_{\mathrm{osc}}}
\newcommand{\Lmod}{L_{\mathrm{mod}}}
\newcommand{\Ltrap}{L_{\mathrm{trap}}}
\newcommand{\Lint}{L_{\mathrm{int}}}
\newcommand{\vl}{v_{\mathrm{L}}}
\newcommand{\al}{\alpha_{\mathrm{L}}}
\newcommand{\vs}{v_{\mathrm{S}}}
\newcommand{\as}{\alpha_{\mathrm{S}}}
\newcommand{\pdt}[1]{\frac{\partial #1}{\partial t}}
\newcommand{\pdz}[1]{\frac{\partial #1}{\partial z}}
\newcommand{\pdzz}[1]{\frac{\partial^2 #1}{\partial z^2}}

\begin{document}


\title{Fermionic shock waves -- dissipative or dispersive?}

\date{\today}

\author{N.K. Lowman}
 \email{nklowman@ncsu.edu}
 \affiliation{Department of Mathematics, North Carolina State
 University, Raleigh, North Carolina 27695, USA}
\author{M. A. Hoefer}
 \affiliation{Department of Mathematics, North Carolina State
 University, Raleigh, North Carolina 27695, USA}


\begin{abstract}
The collision of two clouds of Fermi gas at unitarity (UFG) has been recently observed to lead to shock waves whose regularization mechanism, dissipative or dispersive, is being debated.  While classical, dissipative shocks, as in gas dynamics, develop a steep, localized shock front that translates at a well-defined speed, dispersively regularized shocks are distinguished by an expanding region of short wavelength oscillations with two speeds, those of the leading and trailing edges.  For typical UFG experimental conditions, the theoretical oscillation length scale is smaller than the resolution of present imaging systems so it is unclear how to determine the shock type from its structure alone.  Two experimental methods to determine the appropriate regularization mechanism are proposed:  measurement of the shock speed and observation of a one-dimensional collision experiment with sufficiently tight radial confinement.  

\end{abstract}

\maketitle


Ultracold, dilute gases provide a universal medium for the study of nonlinear hydrodynamic flows in the presence of dissipation and dispersion \cite{kulkarni_hydrodynamics_2012}.  Experimental attainment of the superfluid regime in a Bose-Einstein condensate (BEC) and a unitary Fermi gas (UFG) has led to the observation of nonlinear coherent structures of fundamental interest, including quantized vortices \cite{matthews_vortices_1999, madison_vortex_2000,zwierlein_vortices_2005}, solitons \cite{burger_dark_1999, strecker_formation_2002, yefsah_heavy_2013}, and shock waves \cite{dutton_observation_2001, hoefer_dispersive_2006,joseph_observation_2011}.  In the absence of regularization, the effects of nonlinearity can lead to self-steepening and gradient catastrophe.  In many physical systems, this steepening is mediated by weak dissipation, often due to the effects of viscosity, which transfers kinetic energy to heat across a narrow transition region, a viscous shock wave (VSW).  An alternative regularization mechanism occurs in dissipationless (conservative) media where self-steepening is balanced by dispersion, in which case gradient catastrophe is resolved into the formation of an expanding oscillatory region called a dispersive shock wave (DSW).  For BEC, a mean field description is well-established in which the hydrodynamic equations are regularized by dispersion (c.f. \cite{kevrekidis_emergent_2008}).  Generation of shock waves in recent UFG collision experiments has fueled interest in the formulation of an analogous theory for fermionic systems.  While a direct, computational approach using density functional theory is available \cite{bulgac_quantum_2012}, two simplified models have been proposed and successfully fitted to the experimental data, one in which the hydrodynamic equations are regularized by dissipation \cite{joseph_observation_2011} and the other by dispersion \cite{ancilotto_shock_2012, salasnich_supersonic_2011}.  Although effort has been made to develop theory incorporating dispersion in the weakly interacting regime \cite{bettelheim_quantum_2012, protopopov_dynamics_2013} and both dispersion and dissipation in the weakly nonlinear regime \cite{kulkarni_hydrodynamics_2012}, large amplitude effects and the relative magnitude of dissipation and dispersion in the strongly interacting regime are still unknown. This raises a fundamental question -- how does one determine the appropriate regularization for a UFG from experiment?  An obvious distinction is the structure of the shock.  A VSW takes the form of a traveling wave, while a DSW is characterized by an expanding collection of large amplitude oscillations.  However, in the case of a UFG, the oscillation scale may be too fine to be imaged \cite{joseph_observation_2011,ancilotto_shock_2012}.  In this article, we show that the type of regularization may alternatively be distinguished by measuring the shock speed; VSW and DSW speeds differ.  We also highlight the parameter regime in which a three-dimensional (3D) experiment is amenable to one-dimensional (1D) simplification and argue that the collision experiments in \cite{joseph_observation_2011} were fully 3D, like those observed in BEC \cite{chang_formation_2008}.  We conclude with another regularization distinguishing experiment, the 1D collision problem where a prominent density bulge is predicted for the dissipative case alone.


At zero temperature, the macroscopic, low energy dynamics of a Fermi superfluid can be described by the, as yet unregularized, equations of irrotational hydrodynamics \cite{giorgini_theory_2008}
\begin{equation} \label{eq: 3D continuity}
  \pdt{n}+{\bf \nabla}\cdot(n{\bf v}) = 0 \ ,
\end{equation}
\begin{equation} \label{eq: 3D momentum}
	m \pdt{(n {\bf v})} + m\nabla \cdot \left( n {\bf v} \otimes {\bf v} \right ) +  \nabla P(n) = - n \nabla U({\bf r})  ,
\end{equation}
where ${\bf r} = (x,y,z)$, $U({\bf r})$ is the trap potential, ${\bf v}({\bf r},t)$ is the velocity field, $m$ is the particle mass, $n({\bf r},t)$ is the density, and the pressure law $P(n) = \xi \frac{\hbar^2}{5m} \left( 3\pi^2 n^{5/2}\right) ^{2/3}$ is a scaled version of $\mu(n) = 5P(n)/(2 n)$, the bulk chemical potential.  The total number of particles $N = \int n \, d\mathbf{r}$ is conserved. The irrotational velocity field is proportional to the gradient of a velocity potential $\phi$, ${\bf v} = \frac{\hbar}{2m} {\bf \nabla \phi}$.  The Bertsch parameter, $\xi$, is a dimensionless, universal constant for which we use $\xi \simeq 0.40$ \citep{giorgini_theory_2008}, though more recent studies have suggested a slightly smaller value (e.g. \cite{carlson_auxiliary-field_2011, forbes_resonantly_2011}).  The particular value will not affect the results of our analysis, only the scalings.  In what follows, we consider the case of a harmonic trap potential $U({\bf r}) = \frac{1}{2}\left(\omega_\perp^2r_\perp^2 + \omega_z^2z^2\right)$, where $r_\perp^2 = x^2+y^2$ and $\omega_\perp$, $\omega_z$ are the transverse and longitudinal trap frequencies, respectively.  It is convenient to introduce the harmonic oscillator lengths $a_{\perp,z} = (\hbar/m\omega_{\perp,z})^{1/2}$.

Conservation laws of the form \eqref{eq: 3D continuity}, \eqref{eq: 3D momentum} are known to admit discontinuous shock solutions which, when interpreted in the vanishing viscosity limit, correspond to a generic dissipative regularization of gradient catastrophe \cite{lax_hyperbolic_1973}.  The shock speed is determined by integrating \eqref{eq: 3D continuity}, \eqref{eq: 3D momentum} across a sharp transition resulting in jump conditions.  While the bulk viscosity of a UFG is negligible, its shear viscosity is not \cite{cao_searching_2011} suggesting that dissipation could be a viable regularization mechanism for the singular hydrodynamics.

In contrast, a dispersive regularization of the hydrodynamic equations, proposed in \cite{salasnich_extended_2008}, uses an extended Thomas-Fermi functional approach.  The first order correction to the hydrodynamic system is the addition of a von Weizsacker-type \citep{von_weizsacker_zur_1935}, dispersive correction term to the righthand side of \eqref{eq: 3D momentum} of the form $\lambda\frac{\hbar^2}{4m} \nabla \cdot \left(\rho \nabla \otimes \nabla \log{\rho} \right)$, where $\lambda$ is a dimensionless parameter with accepted value $\lambda \simeq 0.25$ \cite{ancilotto_shock_2012}.  Note that studies in the weakly interacting regime have led to alternative dispersive models  \cite{bettelheim_quantum_2012, protopopov_dynamics_2013}.  While the particular form of the dispersion at the microscopic, oscillatory level is still unknown, the qualitative behaviors due to dispersion which distinguish such systems from their dissipative analogs hold for a broad class of forms (see e.g. \cite{hoefer_shock_2013-1}).

By introducing the complex wavefunction $\psi = \sqrt{n} \exp \left[ i  \phi/(2\sqrt{\lambda})\right]$, the system \eqref{eq: 3D continuity}, \eqref{eq: 3D momentum} with the gradient correction term can equivalently be written in the form of a generalized nonlinear Schr\"{o}dinger (gNLS) equation, similar to the Gross-Pitaevskii equation from BEC mean field theory but with a different nonlinear exponent
\begin{equation} \label{eq: 3D wavefunction}
	i \hbar \lambda^{1/2} \pdt{\psi} = 
	U({\bf r})\psi - \lambda \frac{\hbar^2}{2m} \nabla^2\psi + 
	\xi \mu_0 \left|\psi\right|^{4/3} \psi \ ,
\end{equation}
where  $\mu_0 = \frac{\hbar^2}{2m}\left(3\pi^2\right)^{2/3}$. 

To formulate the 1D shock problem for a UFG, we consider the case of a cigar-shaped trap $\omega_\perp \gg \omega_z$ and derive an effective 1D equation from \eqref{eq: 3D wavefunction}.  We assume sufficiently tight radial confinement so that radial dynamics are negligible and integrate over the transverse coordinates following a standard procedure \cite{menotti_collective_2002,supplementary}.  The anisotropy requirement for dimensionality reduction is $\frac{\omega_z}{\omega_\perp} \ll \frac{1}{N}$ \cite{giorgini_theory_2008}.  This regime is realizable, for example by means of a 2D optical lattice \cite{moritz_confinement_2005, bloch_many-body_2008}.  The 1D wavefunction $\Phi(z,t)$ then satisfies
\begin{equation}	\label{eq: 1D_wavefunction}
  \begin{split}
    i\hbar\lambda^{1/2} \pdt{\Phi}  = &\frac{1}{2}m\omega_z^2 z^2 \Phi   - \lambda\frac{\hbar^2}{2m} \frac{\partial^2 \Phi}{\partial z^2} \\
    &+ \left [ \frac{5m\omega_\perp^2(\xi \mu_0)^{3/2}}{4\pi} \right]^{2/5}\left| \Phi \right|^{4/5} \Phi \ ,
\end{split}
\end{equation}
where $\int |\Phi|^2 dz = N$.
Note that tight transverse confinement leads to a nonlinear coefficient and power that differ from their 3D analog in \eqref{eq: 3D wavefunction}.  It is important to distinguish the various length and time scales in the problem so we nondimensionalize \cite{supplementary} by introducing $\Phi = \Phi_0 \tilde{\Phi}$, $z = L\tilde{z}$, and $t = T\tilde{t}$ with $L=L_0$, $\Phi_0 = \sqrt{N/L_0}$, $T \approx 1.56 [L_0^6/(N \omega_\perp^5 a_\perp^6)]^{1/5}$.  All stated approximate numerical values are given in full in \cite{supplementary}.
$L_0$ is a characteristic length to be chosen.  To obtain the hydrodynamic equations, the longitudinal wavefunction takes the form $\tilde{\Phi} = \sqrt{\tilde{\rho}} \exp \left ( \frac{i}{\epsilon} \int \tilde{u} \, d\tilde{z} \right )$, and after dropping $\tilde{\ }$, the system of 1D conservation equations for UFG is
\begin{equation} \label{eq: mass}
		\pdt{\rho} + \pdz{(\rho u)} = 0,
\end{equation}
\begin{equation} \label{eq: momentum}
	\pdt{ \left( \rho u\right)} + \pdz{ \ }\left( \frac{5}{7}\rho^\frac{7}{5} + \rho u^2 \right) =  \frac{\epsilon^2}{4} \pdz{ \ }\left[ \rho \ \frac{\partial^2  \left( \log{\rho} \right) }{\partial z^2} \right] - \kappa \rho z \ ,
\end{equation}
with
\begin{align}
  \label{eq:2}
  \epsilon &  
  \approx 0.78 \left ( \frac{a_\perp^4}{N L_0^4} \right )^{1/5} \ , \quad
  \kappa 
  \approx 2.45 \left ( \frac{a_\perp^2 L_0^3}{N^{1/2} a_z^5} \right )^{4/5} \ ,
\end{align}
and $\rho$, $u$ are the dimensionless density and longitudinal velocity, respectively.  These hydrodynamic equations admit the long wave speed of sound $c(\rho) = \rho^{1/5}$.  Note that the transformation from \eqref{eq: 1D_wavefunction} to \eqref{eq: mass}, \eqref{eq: momentum} is exact with no approximation.  This form reveals the dispersive regularization of the hydrodynamic equations when $0 < \epsilon \ll 1$.

It is beneficial to briefly describe the relations between inherent length scales associated with different choices for $L_0$.  The longitudinal oscillation length, $\Losc$, associated with DSWs, is obtained by choosing $L_0$ in \eqref{eq:2} so that $\epsilon = 1$.  The longitudinal extent of the trapped UFG, $\Ltrap$, is determined by fixing $L_0$ in \eqref{eq:2} so that $\kappa = 1$.  Another important length scale is associated with the inter-particle spacing $\Lint$ that is estimated by standard arguments \cite{supplementary}.  The 1D anisotropy requirement leads to the following relations
\begin{equation}
  \label{eq:1}
  \Lint \sim \frac{a_\perp^{2/3} a_z^{1/3}}{N^{1/6}} \ll \Losc \sim \frac{a_\perp}{N^{1/4}} \ll \Ltrap \sim N^{1/6} \frac{a_z^{5/3}}{a_\perp^{2/3}} .
\end{equation}
That DSW dynamics occur at length scales much larger than $L_{\mathrm{int}}$ is a requirement for the validity of the hydrodynamic model \citep{giorgini_theory_2008}.  DSWs exhibit rapid oscillations with wavelength $\Losc$ and a larger, envelope modulation length we denote $\Lmod$.  When $L_0 = \Lmod$ such that $\Losc \ll \Lmod \ll \Ltrap$, then $0 < \epsilon \ll 1$ and $0 < \kappa \ll \epsilon^2$ so that a dispersive regularization is appropriate and inhomogeneities due to the trap can be neglected.  In what follows, we use parameters from experiment and set $L_0 = \Lmod = 3 \mu$m corresponding to the experimental imaging resolution \cite{joseph_observation_2011}.  Then, $\epsilon \approx 0.05$ and $\kappa \approx 0.0003$, thus satisfying our smallness conditions.  For the rest of this work we neglect the trap, setting $\kappa = 0$ in \eqref{eq: momentum}.


We now discuss the dynamics of shock solutions for eqs.~\eqref{eq: mass}, \eqref{eq: momentum} with $\epsilon = 0$ (dissipative regularization) and $0<\epsilon\ll 1$ (dispersive regularization).  We consider the conservation laws with general step initial data
\begin{equation}
  \label{eq:4}
	\rho(z,0) = \begin{cases} \rho_0, & z\le 0 \\ 1, & z>0 \end{cases}, \quad u(z,0) = \begin{cases} u_0, & z\le 0 \\ 0, & z>0 \end{cases} .
\end{equation}
For general $\rho_0$ and $u_0$, single step initial data results in the generation of two waves.  We focus on the case $\rho_0 > 1$ and $u_0$ specifically chosen from a locus of velocities corresponding to a single, right propagating shock wave.


In dissipatively regularized systems, shock waves are nonlinear traveling wave solutions which can be viewed as a balance between nonlinearity and dissipation.  Deriving the shock profile for the hydrodynamic equations of UFG with a vanishing effective Newtonian viscosity is mathematically equivalent to constructing a weak solution for the system \eqref{eq: mass}, \eqref{eq: momentum} with $\epsilon=0$.  The magnitude of the dissipative correction sets the width of the localized transition region but does not affect the shock speed.  In order to satisfy the jump conditions for a single right-propagating shock wave, it is required that the velocity of the left state depend on the density $\rho_0$ according to the Hugoniot locus relation \cite{lax_hyperbolic_1973} $u_0(\rho_0) = [5(\rho_0^{7/5}-1)(\rho_0-1)/(7\rho_0)]^{1/2}$. The speed of the corresponding shock is then
\begin{equation} \label{eq:vsw}
	V(\rho_0) =  \{ [5 ( \rho_0^{7/5} - 1) \rho_0] / [ 7(\rho_0-1)] \}^{1/2} \ .
\end{equation}
The VSW speed is parameterized only by the left-state density $\rho_0$ and is independent of the magnitude of the viscous correction.

\begin{figure} \centering
	\includegraphics{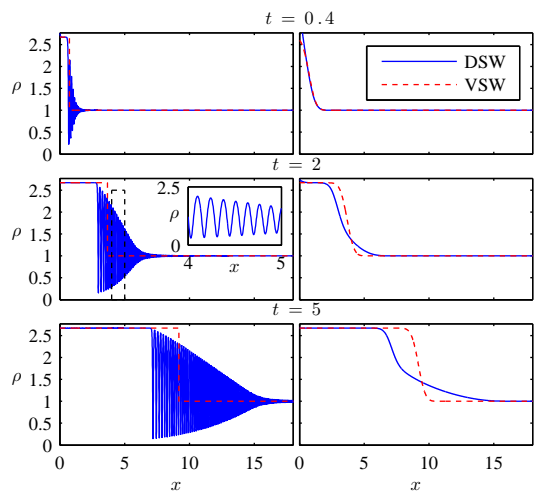}
	\caption{(Color online) Numerical simulation of the gNLS equation (solid) (from
      eqs.~\eqref{eq: mass}, \eqref{eq: momentum}) for a DSW with $\rho_0 = 2.7$ and corresponding VSW solution (dashed).  The
      right panel depicts the filtered solutions.}
\label{fig:averaged_dsw}
\end{figure}

The dispersive regularization, $0<\epsilon\ll 1$, of the initial jump \eqref{eq:4} results in a DSW, two distinct constant states connected by an expanding, oscillatory region.  This region is characterized by nearly linear, vanishing amplitude waves at the rightmost leading edge and a single (dark) solitary wave at the leftmost trailing edge (see Fig.~\ref{fig:averaged_dsw}).  DSW closure conditions, analogous to the jump conditions for VSWs, provide relations between the left and right constant states and the edge speeds.  To implement the DSW closure, we use Whitham-El DSW theory \cite{el_resolution_2005}, an extension of the nonlinear wave modulation theory of Whitham \cite{whitham_non-linear_1965} for dispersive shock--fitting \cite{gurevich_nonstationary_1974} to nonintegrable systems such as the one considered here.  For the theoretical details, we refer the reader to an extensive development of DSW theory for gNLS equations in \cite{hoefer_shock_2013-1}. 

\begin{figure}  \centering 
	\includegraphics{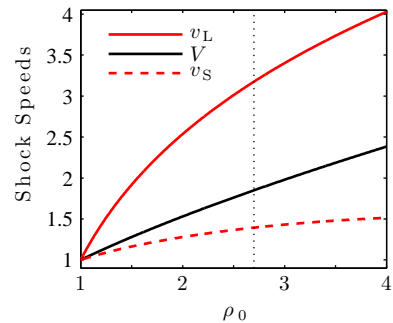}
	\caption{(Color online) Nondimensional shock speeds by type of regularization.  $\vl$ and $\vs$ correspond to the leading and trailing edge speeds of a UFG DSW, respectively.  $V$ is the dissipative shock speed.  The vertical dotted line at $\rho_0 = 2.7$ is for comparison with the particular discontinuity simulated in figure \ref{fig:averaged_dsw}.}
    	\label{fig:speeds}
\end{figure}

DSW closure implies that the velocity $u_0$ must lie on the DSW locus $u_0(\rho_0) = 5( \rho_0^{1/5}-1)$, which is distinct from the Hugoniot locus for VSWs.  The speed $\vl$ of the linear, leading edge satisfies
\begin{equation} \label{eq:dswl}
  \vl = \frac{2\al^2-1}{\al}, \quad
  \frac{27(1+\al)}{2(2+\al)^3} = \rho_0^{-2/5} \ . 
\end{equation}
The speed $\vs$ at the trailing, soliton edge is
\begin{equation} \label{eq:dsws}
  \vs = [ \as\rho_0^{1/5}+5(\rho_0^{1/5}-1)], \quad
  \frac{27(1+\as)}{2(2+\as)^3} = \rho_0^{2/5} \ . 
\end{equation}
While for classical shocks, eq.~\eqref{eq:vsw} gives an explicit speed in terms $\rho_0$, eqs.~\eqref{eq:dswl}, \eqref{eq:dsws} give implicit relations.  To compare them directly, we consider the small jump case $0 < \delta = (\rho_0-1) \ll 1$, in which the dispersively regularized conservation laws admit a weak DSW with speeds $\vl \sim 1+12\delta/5$, $\vs \sim 1+2\delta/5$.
For the weak VSW from \eqref{eq:vsw}, $V \sim 1+ 3\delta/5$.  This analysis demonstrates that regularization dependent shock speeds differ, even for small density perturbations.  In Fig.~\ref{fig:speeds}, we plot the dispersive and viscous shock speeds for a range of density jumps.   As noted earlier, not only the shock speed but also its structure changes drastically depending on the regularization mechanism, though under present imaging capabilities, the difference is not as clear.  The left panel of Fig.~\ref{fig:averaged_dsw} depicts the numerical solution to the gNLS equation corresponding to the hydrodynamic eqs.~\eqref{eq: mass}, \eqref{eq: momentum} and shock solution for the dissipative hydrodynamics $(\epsilon=0)$.  The signature slowly modulated, oscillatory envelope structure expected for a DSW is further revealed by the zoomed-in inset.  The right panels of Fig.~\ref{fig:averaged_dsw} depict the DSW and VSW solutions convolved with a Gaussian of width 1, the modulation length, mimicking the effects of imaging.  Owing to the smoothing of the rapid DSW oscillations, the DSW is now more difficult to distinguish from the filtered VSW solution (dashed).  However, DSW expansion is noticeable from the spreading of the steep density gradient.  Conversely, the VSW maintains its shape and propagates with a different speed.  This shows that determination of the speed of the shock front and its scaling according to \eqref{eq:vsw} or \eqref{eq:dswl},\eqref{eq:dsws} is a viable experimental method for distinguishing between dispersive and dissipative shocks even when potential oscillations are sub-imaging resolution.

\begin{figure} \centering
	\includegraphics{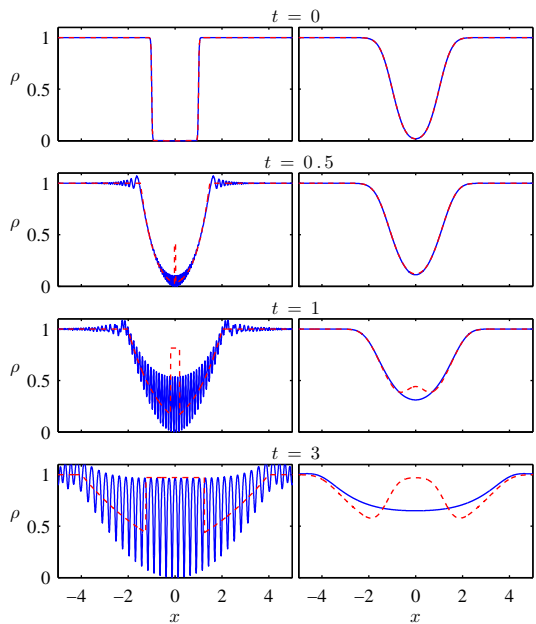}
	\caption{(Color online) Numerical simulation of the gNLS equation (solid) and dissipative hydrodynamic equations (dashed) for the collision
      problem.  The right panel depicts the filtered solutions.}
\label{fig:merging}
\end{figure}

In deriving the shock solutions, we have assumed the validity of the one-dimensional reduction.  In previous attempts to describe the UFG collision experimental observations, a regularization term has been added to a 1D or 3D hydrodynamic model \cite{joseph_observation_2011, ancilotto_shock_2012, salasnich_supersonic_2011}.  From the experimental parameters, it is verified that the anisotropy criterion for reduction to 1D is not met \cite{joseph_observation_2011}.  Further, from its similarity to observations in BEC \cite{chang_formation_2008}, the appearance of a characteristic center ``bulge'' upon collision in the experiments of \cite{joseph_observation_2011} suggests the presence of transverse instabilities and vortex formation, lending doubt to the accuracy of the azimuthally symmetric numerical simulations in \cite{ancilotto_shock_2012}.  The 1D regime, however, is not subject to these instabilities.  Therefore, we consider the 1D collision problem with dissipative ($\epsilon=0$) and dispersive ($0<\epsilon\ll 1$) regularizations for the idealized initial condition
\begin{equation} \label{eq:merge}
\rho(z,0) = \begin{cases} 1, &  \lvert z \rvert > L \\ 0, & \lvert z \rvert \le L \end{cases}, \quad 
u(z,0) = 0 \ ,
\end{equation}
representing two initially separated, quiescent UFGs.  

The results for $L = 1$ are depicted in Fig.~\ref{fig:merging}.  The left panel shows the numerical solution for gNLS and the solution for the corresponding dissipative problem, while the right panel shows the Gaussian filtered solutions.  Initially, two rarefaction waves propagate inward, for which the two regularizations agree to leading order because the effects of dissipation and dispersion are negligible.  Upon collision of the rarefaction waves, outward propagating shocks are created.  In the case of the viscous problem, we derive the weak solution to the hydrodynamic equations \cite{supplementary}.  The dispersive shock solution is similar to that observed in BEC theoretical and experimental studies \cite{hoefer_matterwave_2009}.  The key physical difference after filtering is the distinct center bulge created in the dissipative problem.  The dispersive solution, on the other hand, forms a smooth, expanding well between the two constant states.  Therefore, the one-dimensional collision problem provides a clear distinction between dissipative and dispersive regularizations.

The similarity between the bulge-producing dissipative collision problem and both the BEC \cite{chang_formation_2008} and UFG \cite{joseph_observation_2011} experiments is striking.  We know that BEC is regularized by dispersion, yet a bulge is still produced.  One possible interpretation is that the proliferation of quantized vortices and quantum turbulence in the fully 3D case yields an effective viscosity for the larger scale dynamics, even though the underlying model is dissipationless.  Similarly, fully 3D experiments in a UFG may not provide a definitive determination of the underlying hydrodynamic regularization mechanism unless imaging resolution is significantly improved.  Our work here provides unambiguous measures of dispersive/dissipative effects in a unitary Fermi gas from shock dynamics, enabling the determination of the dominant regularization mechanism.



\emph{We thank John Thomas, James Joseph, and Peter Engels for enlightening discussion.  This work was supported by NSF Grant Nos. DGE--1252376 and DMS--1008973.}


\section{Appendix}

The 3D wavefunction expression (3) was reduced to an effective 1D model by integrating over the transverse dynamics as described in \citet{menotti_collective_2002}.  In this appendix we present the details of this calculation, as well as the derivation of the characteristic length and time scales presented.  We further derive the weak solution for the dissipative 1D collision problem.  We assume that the anisotropy condition is satisfied.

Decompose the 3D complex wavefunction such that $\psi({\bf r},t) = \Psi({\bf r_\perp};n_1(z,t))\Phi(z,t)$, where ${\bf r} = ({\bf r_\perp}, z)$ and $n_1(z,t)$ is the local density function which satisfies the normalization condition
\begin{equation} \label{eq:normalization}
	n_1(z,t) = N \int \left| \psi({\bf r},t) \right|^2 d{\bf r}	 \ .
\end{equation}
Also assume the trap potential takes the special form $U({\bf r}) = U_1({\bf r_\perp}) + U_2(z)$ (above we consider the case of an ideal harmonic trap which certainly fits into this assumption).  Normalize the transverse and longitudinal wave functions according to the constraints
\begin{equation} \label{eq: normalization}
	\int \left| \Psi({\bf r_\perp};n_1)\right|^2 d{\bf r_\perp} = 1, \quad \int \left| \Phi(z,t) \right|^2 dz = N  \ .
\end{equation}	
Hence, $n_1(z,t) = N\left|\Phi(z,t)\right|^2$.  Substituting these anzatzes into (3) yields 
\begin{align}	\label{eq: star}
	& \left[  i\hbar \lambda^{1/2} \pdt{\Phi} +\lambda\frac{\hbar^2}{2m} \frac{\partial^2 \Phi}{\partial z^2} - U_2(z)\Phi\right] \Psi = \\ 
	& \quad \left[-\lambda\frac{\hbar^2}{2m}\Delta_r \Psi + U_1({\bf r_\perp}) \Psi +  \xi\mu_0\left(\left| \Psi \Phi \right|^{4/3}  \right) \Psi \right] \Phi	 \ .
\end{align}
Multiply the equation by the complex conjugate $\Psi^*$ and integrate with respect to the transverse coordinates ${\bf r_\perp}$.  Using the normalization condition on $\Psi$, gives
\begin{equation} \label{eq: 1D}
	 i\hbar \lambda^{1/2} \pdt{\Phi} +\lambda\frac{\hbar^2}{2m} \pdzz{\Phi} - U_2(z)\Phi = \mu_\perp(n_1)\Phi,
\end{equation}
where
\begin{equation}	\label{eq: muperp}
	\begin{split}
	& \mu_\perp(n_1) = \\
	& \quad \int \Psi^* \left[-\lambda\frac{\hbar^2}{2m}\Delta_r  + U_1({\bf r_\perp})  +  \xi\mu_0 \left(\left| \Psi\Phi \right|^{4/3} \right) \right] \Psi \ d{\bf r_\perp}  \ .
	\end{split}
\end{equation}
This is now an effectively 1D equation, but it is left to determine $\mu_\perp$ and $\Psi$.  If we substitute the expression for $\mu_\perp$ \eqref{eq: muperp} into eq.~\eqref{eq: star} and enter the Thomas-Fermi regime (neglecting the radial Laplacian term), we get the following eigenvalue problem for $\Psi$:
\begin{equation}
	 U_1({\bf r_\perp}) \Psi +  \xi\mu_0\left(\left| \Psi\Phi \right|^{4/3}\Psi  \right)  = \mu_\perp(n_1)\Psi  \ .
\end{equation}
Note that $\left| \Phi \right|^{4/3} = \left( n_1 / N \right)^{2/3}$.  Then solving for $\Psi$, we find
\begin{equation} \label{eq:psi}
	\left|\Psi\right|^{4/3} = \left( \frac{N}{n_1} \right)^{2/3}\frac{1}{\xi\mu_0}  \left( \mu_\perp - U_1({\bf r_\perp})\right)  \ .
\end{equation}
Assuming that $\Psi$ is real in the ground state, \eqref{eq:psi} can be solved explicitly to obtain
\begin{equation} \Psi = \begin{cases} \displaystyle
	\left( \frac{N}{n_1} \right)^{1/2} \left[\xi\mu_0\left( \mu_\perp - U_1({\bf r_\perp})\right)\right]^\frac{3}{4}, & \mu_\perp > U_1({\bf r_\perp}), \\
	0, & \text{otherwise}  \ .
	\end{cases}	\label{eq: transverse}
\end{equation}
Next, an expression is needed for $\mu_\perp$ which comes from the normalization condition on the transverse wave function \eqref{eq: normalization}.  Assume an ideal, harmonic trap potential so that $U_1({\bf r_\perp}) = \frac{1}{2} \omega_\perp^2 r_\perp^2$, where $r_\perp^2 = x^2 + y^2$.  Note that $\Psi$ vanishes for $r_\perp$ such that $U_1({\bf r_\perp}) \ge \mu_\perp$.  This gives the upper bound of the integration in \eqref{eq:normalization} to be
\begin{equation}	\label{eq: r_cond}
	r_\perp < \left( \frac{2\mu_\perp}{m\omega_\perp} \right) ^{1/2}	 \ .
\end{equation}
Substituting eqs.~\eqref{eq: transverse} and \eqref{eq: r_cond} into eq.~\eqref{eq: normalization}, one can solve for $\mu_\perp$ from
\begin{equation} \displaystyle \label{eq: mu_int}
	\int_0^{\left( \frac{2\mu_\perp}{m\omega_\perp} \right) ^{1/2}} r_\perp \left( \mu_\perp - \frac{1}{2} \omega_\perp^2 r_\perp^2 \right)^{3/2} dr_\perp = \frac{(\xi\mu_0)^{3/2}}{2\pi} \left( \frac{n_1}{N} \right)  \ .
\end{equation} 
Eq.~\eqref{eq: mu_int} can be integrated upon making a simple change of variables to obtain
\begin{equation}
	\mu_\perp = \left( \frac{5m\omega_\perp^2(\xi\mu_0)^{3/2}}{4\pi} \right)^{2/5}\left| \Phi \right|^{4/5}  \ .
\end{equation}
The trap potential in the longitudinal coordinate $z$ is assumed to be $U_2(z) = \frac{1}{2}m\omega_z^2 z^2$.  Hence, eq.~\eqref{eq: 1D} becomes
\begin{equation}	\label{eq: 1D_2}
  \begin{split}
	&i\hbar\lambda^{1/2} \pdt{\Phi} +\lambda\frac{\hbar^2}{2m} \pdzz{\Phi}- \frac{1}{2}m\omega_z^2 z^2\Phi -  \\
	&\quad \left( \frac{5m\omega_\perp^2(\xi\mu_0)^{3/2}}{4\pi} \right)^{2/5}\left| \Phi \right|^{4/5} \Phi=0,
  \end{split}
\end{equation}
which is 1D and in dimensional form.  

We now nondimensionalize the equation using $\Phi = \Phi_0 \tilde{\Phi}$, $z = L\tilde{z}$, and $t = T\tilde{t}$ to the semiclassical scaling with normalized sound speed, which after dropping tildes becomes
\begin{equation} \label{eq: semiclassical}
	i\epsilon \pdt{\Phi} + \frac{\epsilon^2}{2}\pdzz{\Phi} - \frac{1}{2} \kappa z^2\Phi - \frac{5}{2}\left|\Phi\right|^{4/5}\Phi=0
	\ .
\end{equation}

Note that each term in \eqref{eq: 1D_2} is dimensional.  We multiply both sides by a dimensional parameter $A$ and normalize \eqref{eq: 1D_2} to the semiclassical scaling \eqref{eq: semiclassical} while enforcing the normalization condition \eqref{eq: normalization}.  This yields the parameters and scalings 
\begin{equation}
	L=L_0, \quad \Phi_0 = \left( \frac{N}{L_0} \right)^{1/2}, 
\end{equation}
\begin{equation}
	T = \left( \frac{10\sqrt{5}}{3\pi\xi^{3/2}N} \right)^{1/5} \left( \frac{1}{\omega_\perp} \right)
	\left( \frac{L_0}{a_\perp} \right)^{6/5}
\end{equation}
\begin{equation}
	\epsilon = \left( \frac{10 \sqrt{5}\lambda^{5/2}}{3\pi\xi^{3/2}N} \right)^{1/5} 
	\left( \frac{a_\perp}{L_0} \right)^{4/5},
\end{equation}
\begin{equation}
	\kappa = \left( \frac{10\sqrt{5}}{3\pi\xi^{3/2}N} \right)^{2/5} \left( \frac{\omega_z}{\omega_\perp} \right)^2 \left( \frac{L_0}{a_\perp} \right)^{12/5},
\end{equation}
\begin{equation}
	A = \left( \frac{10\sqrt{5}}{3\pi\xi^{3/2}N} \right)^{2/5} \left( \frac{1}{\hbar\omega_\perp} \right)
	\left( \frac{L_0}{a_\perp} \right)^{2/5} \ .
\end{equation}

The inter-particle spacing $\Lint$ is estimated by determining the
approximate volume occupied by $N$ Fermions in the ground state of the
harmonic trap.  The standard Thomas-Fermi approximation is used
whereby the kinetic energy terms are neglected in favor of a balance
between the nonlinearity and the trap potential, $\mu(n) = \mu_0 -
U(\mathbf{r})$ from eq.~(2).  $\mu_0$ is the total chemical potential determined by
the requirement that $\int n \, d\mathbf{r} = N$.  The volume is
\begin{equation}
  \label{eq:5}
  \begin{split}
    \mathrm{Vol} \{\mathbf{r} \, | \, \mu_0 > U(\mathbf{r})\} &= \frac{2^{7/2} \xi^{3/4} \pi}{3^{1/2}} N^{1/2} a_\perp^2 a_z \\
    &\approx 10.3 \,N^{1/2} a_\perp^2 a_z .
  \end{split}
\end{equation}
Dividing the cube root of the volume by $N$ gives
\begin{equation}
  \label{eq:3}
  \Lint = \left ( \frac{2^{7/2} \xi^{3/4} \pi}{3^{1/2}} \right )^{1/3}
  \left ( \frac{a_\perp^2 a_z}{N^{1/2}} \right )^{1/3} \approx 2.18
  \left ( \frac{a_\perp^2 a_z}{N^{1/2}} \right )^{1/3} .
\end{equation}

The 1D collision problem with $\epsilon=0$ and initial data given by (13) generates two rarefaction waves propagating inward from the initial discontinuities at $\lvert z \lvert = L$, which interact at time $t=t_\mathrm{i}$.  For comparison with numerical simulations, we set $L=1$, though the results are easily generalizable.  On the right (the left can be constructed from symmetry), the rarefaction wave is given by 
\begin{equation} \displaystyle
	u \left(z,t \right) = \frac{5}{6} \left[\frac{z-1}{t} - 1\right] \ ,
\end{equation}
\begin{equation} \displaystyle
	\rho\left(z,t \right) = \left[ \frac{1}{6} \left( 5 + \frac{z-1}{t}\right) \right]^5 \ ,
\end{equation}
for $(z-1)/t \in [-5, 1]$.  Hence, $t_\mathrm{i} = 1/5$.  Immediately upon interaction, an intermediate velocity state $u_\mathrm{m} = 0$ is created along the center axis $z=0$ and two outward propagating shocks are produced.  These can be described by constructing an intermediate density $\rho_\mathrm{m}$ so that it lies along the appropriate density locus, i.e.
\begin{equation}	\label{eq:locus}
	u^2 = \frac{5\left(\rho_\mathrm{m}^{7/5} - \rho^{7/5}\right)\left(\rho_\mathrm{m} - \rho\right)}
		{7 \rho_\mathrm{m}\rho} \ ,
\end{equation}
where $u$ and $\rho$ are evaluated at the right rarefaction waves just ahead of the shock front, which are the variable background into which the shock is propagating.  The intermediate value $\rho_\mathrm{m}$, which must be found by evaluating \eqref{eq:locus} implicitly, couples to the shock speed by invoking the jump conditions to give the ordinary differential equation 
\begin{equation}
  \label{eq:6}
	s'(t) = \left. \frac{\rho u}{\rho - \rho_\mathrm{m}} \right|_{z=s(t)} \ .
\end{equation}
The initial condition must be prescribed just after the interaction time so that a shock is created, say $t = t_\mathrm{i} + \nu$, $0<\nu \ll 1$.  Then, inserting the Taylor series expansion, $s(1/5)=0\sim s(1/5+\nu) - \nu s'(1/5+\nu)$ into \eqref{eq:locus}, \eqref{eq:6} gives the approximate initial data
\begin{equation}
	s(1/5 + \nu) \sim \nu^{17/7}\frac{5^{17/7}}{35^{5/7} - 1}, \quad 0 < \nu \ll 1 \ ,
\end{equation}
which we use as initial conditions to numerically solve the system \eqref{eq:locus}, \eqref{eq:6}.  For the simulations presented, we took $\nu = 5\times 10^{-5}$ and found it to be sufficiently small to accurately resolve the shock dynamics.


\bibliographystyle{apsrev4-1}

\end{document}